\documentclass{ws-rv975x65}
\usepackage{subfigure}     
\usepackage{ws-rv-thm}     
\usepackage{ws-rv-van}     
\makeindex

\newcommand{\be}{\begin{equation}}
\newcommand{\ee}{\end{equation}}
\newcommand{\ba}{\begin{array}}
\newcommand{\ea}{\end{array}}
\newcommand{\bea}{\begin{eqnarray}}
\newcommand{\eea}{\end{eqnarray}}

\newcommand{\cO}{{\cal O}}
\newcommand{\cB}{{\cal B}}

\newcommand{\cL}{{\cal L}}

\newcommand{\cM}{{\cal M}}
\newcommand{\cG}{{\cal G}}

\newcommand{\no}{\nonumber}

\newcommand{\sla}{\! \! \! \!  /~}
\newcommand{\rhob}{\bar\varrho}
\newcommand{\etab}{\bar\eta}

\newcommand{\gsim}{\lower.7ex\hbox{$\;\stackrel{\textstyle>}{\sim}\;$}}
\newcommand{\lsim}{\lower.7ex\hbox{$\;\stackrel{\textstyle<}{\sim}\;$}}

\begin{document}

\chapter[Flavour Physics and Implication for New Phenomena]{Flavour Physics and Implication for New Phenomena}

\author[Gino Isidori]{Gino Isidori} 

\address{Physik-Institut, Universit\"at Z\"urich, CH-8057 Z\"urich, Switzerland,   \\ 
INFN, Laboratori Nazionali di Frascati, I-00044 Frascati, Italy}  

\begin{abstract}
Flavour physics represents one of the most interesting and, at the same time, less understood sector of the Standard Theory.
On the one hand, the peculiar pattern of quark and lepton masses, and their mixing angles, may be the clue to some new dynamics 
occurring at high-energy scales. On the other hand, the strong suppression of flavour-changing neutral-current processes, predicted by the Standard  Theory and confirmed by experiments, represents a serious challenge to extend the Theory. 
This article reviews both these aspects of flavour physics from a theoretical perspective.
\end{abstract}
\body

\section{Introduction}

The term {\em flavour} is used, in the jargon of particle physics, to characterize 
the different copies of fields with the same spin and gauge quantum numbers,
and {\em flavour physics} refers to the study of the interactions that distinguish between these copies. 
Within the Standard Theory (ST) of fundamental interactions, as we know it now, all matter fields (quark, leptons, 
and neutrinos) appear in three flavours, and the only interaction 
that distinguish these three flavours is the Yukawa
interaction, or the interaction of the matter fields 
with the Higgs boson. 

The fact that flavour non-universality is generated only by Yukawa interaction 
is an unavoidable consequence in the Standard Theory, given its  particle content. 
However, this structure was  far from being obvious for decades: from the discovery of strange particles in the 1950's till the triumph 
of the ST predictions for quark-flavour mixing  observed at the $B$-factories in the 2000's. 
During all these years the progress in understanding flavour physics has been intimately related to the overall
progress in building and testing the ST of fundamental interactions.

At present we have a clear understanding of the underlying mechanism of flavour mixing and flavour non-universality within the ST, 
and this mechanism has been successfully verified in experiments. However,  flavour physics still represents one of the most puzzling and, at the same time, interesting 
aspects of particle physics.  Our ``ignorance'' in this sector can be summarized by the following two open questions:
\begin{itemize}
\item{} What determines the observed pattern of masses and mixing angles of quarks
and leptons?
\item{} Which are the sources of flavour symmetry breaking accessible at low energies?
Is there anything else beside the ST Yukawa couplings?
\end{itemize}
Answering these questions is a key part of the more general program of  investigating  
the nature of  physics beyond the ST. There are indeed convincing arguments, including the peculiar  pattern of quark 
and lepton masses,
which motivate  us to consider the ST as the low-energy limit of a more complete theory.

The precise understanding of  the mechanism of flavour mixing within the ST, summarized in section 2-3,
is essential to formulate the above questions in a quantitative way. The present status 
of the  partial answers obtained so far to the second question, and their implications for physics beyond the ST,
are presented in section 4-6. Some of the theoretical ideas put forwards to 
address the first question are presented in section 7.  

\section{Some historical remarks}

The first building block of what 
we now call flavour physics was laid down by Cabibbo in 
1963~\cite{Cabibbo:1963yz}, well before many of the ingredients
of the Standard Theory were clear. The Cabibbo theory 
of semileptonic decays  provided the first step toward a unified 
description of hadronic and leptonic weak interactions.
Later on, the hypothesis of the existence of the charm quark, 
formulated by Glashow, Iliopoulos and Maiani~\cite{Glashow:1970gm}, represented 
a key ingredient both to understand the mechanism of 
quark flavour mixing within the ST and, at the same time,
to understand how to extend the unified mechanism 
of weak and electromagnetic interactions from the lepton
sector to the quark sector. Finally, the hypothesis formulated 
by Kobayashi and Maskawa~\cite{Kobayashi:1973fv} that quarks 
appear in three flavour turned out to be the correct explanation 
of the phenomenon of CP violation within the ST. 

The theoretical foundations of the mechanism of flavour mixing within the ST
were anticipated and followed by a long series of key experimental observations, 
starting from the discovery of CP violation in the neutral kaon system
in 1964~\cite{Christenson:1964fg},  and culminated with the precise determination 
of all the elements of the Cabibbo-Kobayashi-Maskawa quark-flavour mixing matrix 
at the $B$-meson factories\cite{Bevan:2014iga},  and at various dedicated $K$-decay 
experiments\cite{Antonelli:2010yf,Cirigliano:2011ny}. 
At the completion of the $B$-factory program, it has became 
clear that the ST provides a successful   description of the mechanism of quark flavour
mixing: possible contributions due to New Physics (NP), if any, can only be small corrections 
compared to the leading ST terms. The search for such tiny deviations is the main 
goal of present and future experimental efforts in  flavour physics.\cite{Bediaga:2012py,Abe:2010gxa,fortheNA62:2013jsa,
Yamanaka:2012yma}

The precise comparison between data and ST  in flavour physics has been 
made possible by a significant amount of theoretical progress in understanding 
how QCD interactions modify weak interactions at low energies. This started with the pioneering work of 
Gaillard and Lee~\cite{Gaillard:1974hs}, and Altarelli and Maiani~\cite{Altarelli:1974exa},
further extended by Shifman,  Vainshtein, and Zakharov\cite{Shifman:1975tn}, 
and by Gilman and M. B. Wise\cite{Gilman:1979bc}.  
A significant step forward was undertaken in the 1990's, where all the relevant flavour-changing processes
have been computed at the next-to-leading order (NLO) accuracy, in particular by the 
Munich~\cite{Buchalla:1993bv} and Rome~\cite{Ciuchini:1993vr} groups (see Ref.~\refcite{Buchalla:1995vs} 
for a complete list of NLO references). More recently specific processes, 
such as $B\to X_s\gamma$ and   $B_{s,d} \to \ell^+ \ell^-$,  
have been computed even at NNLO accuracy~\cite{Misiak:2015xwa,Czakon:2015exa,Bobeth:2013uxa}.

\section{The  flavour sector of the Standard Theory}
\label{sect:ST}

The  ST Lagrangian can be divided into two main parts, 
the gauge and the Higgs (or symmetry breaking) sector. The gauge sector
is extremely simple and highly symmetric: it is completely specified by the 
local symmetry $SU(3)_{C}\times SU(2)_{L}\times U(1)_{Y}$
and by the fermion content. This consists of five fields with different quantum numbers 
under the gauge group: the $SU(2)_L$ doublet of quarks ($Q^i_{L}$), 
the two right-handed quark singlets ($U^i_{R}$ and $D^i_{R}$ ), 
the lepton doublet ($Q^i_{L}$), and the 
right-handed lepton singlet ($E^i_{R}$).

Each of these five different fields appears in three different replica or 
flavours ($i=1,2,3$), giving rise to a large {\em global} flavour symmetry. 
Both the local and the global symmetries of the  gauge sector of the ST 
are broken by the Higgs field. The local symmetry is spontaneously 
broken by the vacuum expectation value of the Higgs field,
$\langle |\phi | \rangle  = v = (2\sqrt{2} G_F)^{-1/2} \approx 174$~GeV, while 
the global flavour symmetry is {\em explicitly broken} by 
the Yukawa interaction of $\phi$ with the fermion fields:
\be
\label{eq:STY}
- {\cal L}^{\rm ST}_{\rm Yukawa}=Y_d^{ij} {\bar Q}^i_{L} \phi D^j_{R}
 +Y_u^{ij} {\bar Q}^i_{L} \tilde\phi U^j_{R} + Y_e^{ij} {\bar L}_{L}^i
\phi E_{R}^j + {\rm h.c.} \qquad  ( \tilde\phi=i\tau_2\phi^\dagger)~.  
\ee
The large global flavour symmetry of  $\cL^{\rm ST}_{\rm gauge}$, 
corresponding to the independent unitary rotations in flavour space 
of the five fermion fields, is a $U(3)^5$ group.\cite{Chivukula:1987py}
This can be decomposed as follows: 
${\mathcal G}_{\rm flavour} = U(1)^5 \times  
{\mathcal G}_{q} \times {\mathcal G}_{\ell}~, 
\label{eq:Gtot}$
where 
\be
{\mathcal G}_{q} = {SU}(3)_{Q_L}\times {SU}(3)_{U_R} \times {SU}(3)_{D_R}, \qquad 
{\mathcal G}_{\ell} =  {SU}(3)_{L_L} \times {SU}(3)_{E_R}~.
\label{eq:Ggroups}
\ee  
Three of the five $U(1)$ subgroups can be identified with the total barion and 
lepton number, which are not broken by the Yukawa interaction, and the weak hypercharge, 
which is gauged and broken only spontaneously by $\langle \phi \rangle  
\not=0$. The subgroups controlling flavour-changing dynamics
and flavour non-universality  are the non-Abelian groups ${\mathcal G}_{q}$ 
and ${\mathcal G}_{\ell}$, which are explicitly broken by $Y_{d,u,e}$ not being 
proportional to the identity matrix. 

The diagonalization of each Yukawa matrix requires, in general, two 
independent unitary matrices, $V_L Y V^\dagger_R = {\rm diag}(y_1,y_2,y_3)$.
In the lepton sector we are free to choose the two matrices 
necessary to diagonalize  $Y_e$ without breaking gauge invariance. This is not the case in the quark 
sector, where we cannot diagonalize on the left both $Y_{d}$ and $Y_u$ at the same time. 
We are thus left with a non-trivial misalignment matrix $V$, between  $Y_{d}$ and $Y_u$, 
which is nothing but the Cabibbo-Kobayashi-Maskawa (CKM) mixing matrix~\cite{Cabibbo:1963yz,Kobayashi:1973fv}:
\be 
V=\left(\begin{array}{ccc}
V_{ud}&V_{us}&V_{ub}\\
V_{cd}&V_{cs}&V_{cb}\\
V_{td}&V_{ts}&V_{tb}
\end{array}\right) 
\ee

For practical purposes it is often convenient to work in the mass eigenstate basis 
of both up- and down-type quarks. This can be achieved rotating independently 
the up and down components of the quark doublet $Q_L$, or moving the CKM matrix 
from the Yukawa sector to the charged weak current in $\cL^{\rm ST}_{\rm gauge}$:
\be
\left. J_W^\mu \right|_{\rm quarks} = \bar u^i_L \gamma^\mu d^i_L \quad \stackrel{u,d~{\rm mass-basis}}{\longrightarrow} \quad
\bar u^i_L V_{ij} \gamma^\mu d^j_L ~.
\label{eq:Wcurrent}
\ee
However, it must be stressed that $V$ originates from the Yukawa sector (in particular 
by the miss-alignment of $Y_u$ and $Y_d$ in the ${SU}(3)_{Q_L}$ subgroup of ${\mathcal G}_q$): 
in the absence of Yukawa  couplings we can always set $V_{ij}=\delta_{ij}$.

To summarize, 
quark flavour physics within the ST is characterized by a large flavour symmetry, 
${\mathcal G}_{q}$, defined by the gauge sector, whose only breaking sources 
are the two Yukawa couplings $Y_{d}$ and $Y_{u}$. The CKM matrix arises by the 
miss-alignment of $Y_u$ and $Y_d$ in flavour space.

\subsection{The CKM matrix}

The residual invariance under the flavour group allows us to eliminate five of the six complex phases in $V$,
that contains only four real physical parameters: three mixing angles and
one  CP-violating phase.  
%
The off-diagonal elements of the CKM matrix
show a strongly 
hierarchical pattern:  $|V_{us}|$ and $|V_{cd}|$ are close to $0.22$, the elements
$|V_{cb}|$ and $|V_{ts}|$ are of order $4\times 10^{-2}$ whereas $|V_{ub}|$ and
$|V_{td}|$ are of $O(10^{-3})$. 

The Wolfenstein parametrization, namely the expansion of the CKM matrix 
elements in powers of the small parameter $\lambda \doteq |V_{us}| \approx 0.22$, is a
convenient way to exhibit this hierarchy in a more explicit way~\cite{Wolfenstein:1983yz}:
\begin{equation}
V=
\left(\begin{array}{ccc}
1-\frac{\lambda^2}{2}&\lambda&A\lambda^3(\varrho-i\eta)\\ -\lambda&
1-\frac{\lambda^2}{ 2}&A\lambda^2\\ A\lambda^3(1-\varrho-i\eta)&-A\lambda^2&
1\end{array}\right)
+{\cal{O}}(\lambda^4)~.
\label{eq:Wolfpar} 
\end{equation}
Here $A$, $\varrho$, and $\eta$ are three independent parameters of order 1. 
Because of the smallness of $\lambda$ and the fact that for each element 
the expansion parameter is actually $\lambda^2$, this is a rapidly converging
expansion. 

\begin{figure}[t]
\begin{center}
\setlength{\unitlength}{1\linewidth}
\includegraphics[width=0.50\linewidth,height=0.48\linewidth]{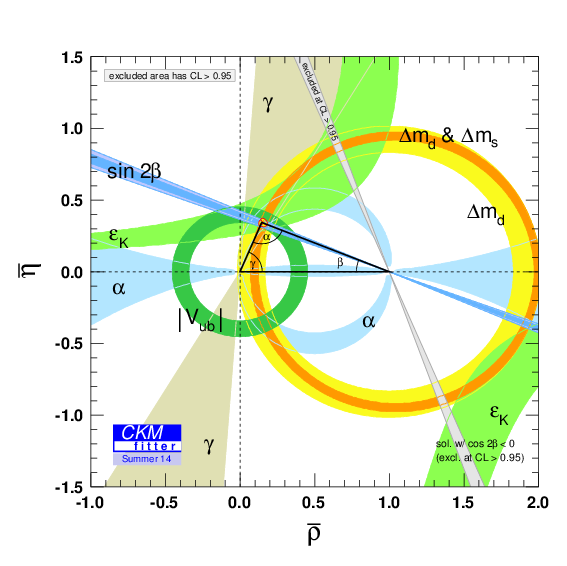}
\includegraphics[width=0.49\linewidth,height=0.48\linewidth]{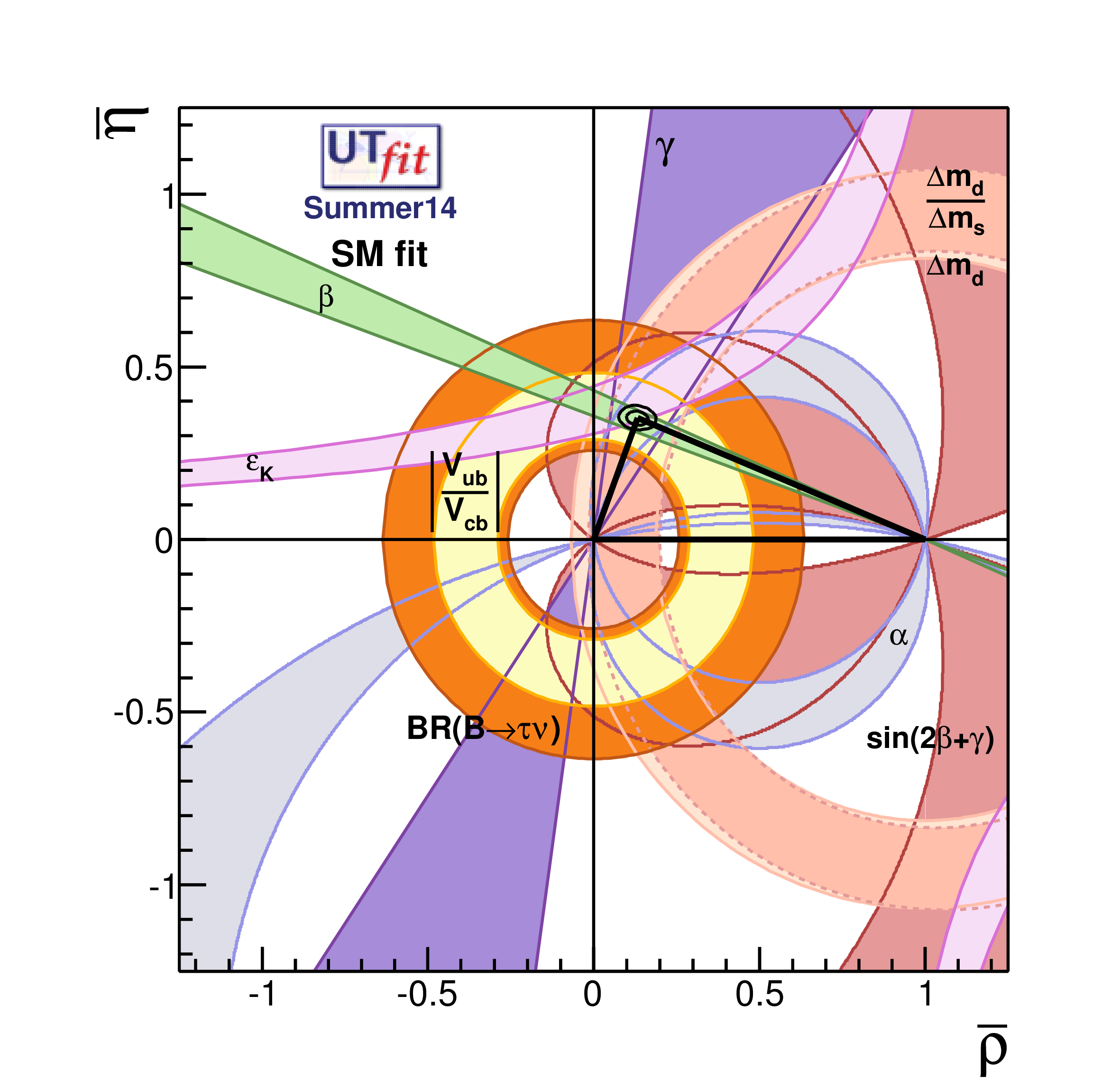}
\end{center}
  \caption{Allowed region in the $\rhob,\etab$ plane as obtained by the CKMfitter~\cite{ckmfitter} and UTfit~\cite{UTfit}
collaborations.  Superimposed are
  the individual constraints from charmless semileptonic $B$ decays
  ($|V_{ub}|$), mass differences in the $B_d$ ($\Delta m_d$)
  and $B_s$ ($\Delta m_s$) systems, CP violation 
  in the neutral kaon ($\varepsilon_K$) and in the $B_d$ systems ($\sin2\beta$),
  the combined constrains on $\alpha$ and $\gamma$ from various $B$ decays. }
  \label{fig:UT}
\end{figure}

The unitarity of the CKM matrix implies a series of relations of the type 
$\sum_{k=1\ldots 3} V_{ki}^* V_{kj}= \delta_{ij}$.
These relations are a distinctive feature of the ST, where the CKM matrix is the only 
source of quark flavour mixing.  Their experimental verification is therefore a useful 
tool to set bounds on, or possibly reveal, new sources of flavour symmetry breaking. 
Among these relations, the one obtained for $i=1$ and $j=3$, 
namely 
\be
V_{ud}^{}V_{ub}^* + V_{cd}^{}V_{cb}^* + V_{td}^{}V_{tb}^* =0 
\label{eq:UT}
\ee
\be
{\rm or} \qquad 
\frac{V_{ud}^{}V_{ub}^*}{V_{cd}^{}V_{cb}^*}  + \frac{V_{td}^{}V_{tb}^*}{V_{cd}^{}V_{cb}^*}  + 1 = 0
\qquad \leftrightarrow \qquad
 [\rho +i \eta ] + [(1-\rho )-i\eta ] + 1 =0~,
\no
\ee
is particularly interesting since it involves the sum of three terms all
of the same order in $\lambda$ and is usually represented as a unitarity triangle
in the complex  plane (see Fig.~\ref{fig:UT}). 
We stress that Eq.~(\ref{eq:UT}) is invariant under any 
phase transformation of the quark fields. Under such transformations
the unitarity triangle  is rotated in the complex plane,
but its angles and the sides remain unchanged.
Both angles and  sides of the unitary triangle are indeed observable quantities
which can be measured in suitable experiments.

The values of $|V_{us}|$ and $|V_{cb}|$ (or $\lambda$ and $A$),
 are determined with good accuracy from  $K\to\pi\ell\nu$ and 
$B\to X_c \ell\nu$ decays, respectively. 
Using these inputs, all the other constraints on the elements of 
the CKM matrix can be expressed as constraints on $\rho$ and $\eta$.
The list of the most sensitive observables used to (over) 
determine  the CKM matrix elements include  (see Fig.~\ref{fig:UT}):
\begin{itemize}
\item The rates of inclusive and exclusive charmless semileptonic $B$
   decays, that depend on $|V_{ub}|$.
\item The phase of the $B_d$--$\bar B_d$ mixing amplitude 
(measured from the time-dependent CP asymmetry in $B\to\psi K_S$ decays),
    that depends on $\sin2\beta$.
\item The rates of various $B\to DK$ decays  constraining the 
  angle  $\gamma$.
\item The rates of various $B\to\pi\pi,\rho\pi,\rho\rho$ decays constraining the 
   combination $\alpha=\pi-\beta-\gamma$.
\item The ratio between the mass splittings in the neutral $B$ and
  $B_s$ systems, that depends on $|V_{td}/V_{ts}|$.
\item The indirect CP violating parameter of the kaon system 
  ($\epsilon_K$), that determines a hyperbola in the $\rho$--$\eta$
  plane.
\end{itemize}
The resulting constraints, as implemented by the CKMfitter and UTfit collaborations,
are shown in Fig.~\ref{fig:UT}.  As can be seen, they are all consistent with 
a unique value of $\rhob= \rho (1-\frac{\lambda^2}{2})$ and 
$\etab=\eta (1-\frac{\lambda^2}{2})$.

The consistency of different constraints 
on the CKM unitarity triangle is a powerful consistency test 
of the ST in describing flavour-changing phenomena.
From the plot in Fig.~\ref{fig:UT} it is quite clear, 
at least in a qualitative way, that there is little room 
for non-ST contributions in flavour changing transitions. 
A more quantitative evaluation of this 
statement is presented in the next section.

\section{The flavour problem}

As anticipated in the introduction, despite the impressive phenomenological 
success of the ST, there are various 
convincing arguments which motivate us to consider this model only as the 
low-energy limit of a more complete theory.

Assuming that the new degrees of freedom which complete the theory 
are heavier than the ST particles,
we can integrate them out and describe physics beyond the ST in full
generality by means of an effective field theory (EFT) approach.
The ST Lagrangian becomes the renormalizable part of a more 
general local Lagrangian which includes an infinite tower of 
operators with dimension $d>4$, constructed in terms of the ST fields
and suppressed by inverse powers of an effective scale 
$\Lambda$. These operators are the residual effect of 
having integrated out the new heavy degrees of freedom, whose
mass scale is parametrized by the effective scale $\Lambda > m_W$. 

Integrating out heavy degrees of freedom is a procedure
often adopted also within the ST: it allows us to 
simplify the evaluation of amplitudes which involve 
different energy scales. This approach is indeed
a generalization of the Fermi theory of weak interactions,
where the dimension-six four-fermion operators describing 
weak decays are the results of having integrated out 
the $W$ field. The only difference when applying this 
procedure to physics beyond the ST is that in this case,
as also in the original work by Fermi, we don't know the 
nature of the degrees of freedom we are integrating out. 
This implies  we are not able to determine a priori the 
values of the effective couplings of the higher-dimensional 
operators. The advantage of this approach is that it 
allows us to analyse all realistic extensions of the ST 
in terms of a limited number of free parameters.

The Lagrangian of the ST considered 
as an effective theory can be written as follows
\be
\cL_{\rm eff} = \cL^{\rm ST}_{\rm gauge} + \cL^{\rm ST}_{\rm Higgs} + 
\cL^{\rm ST}_{\rm Yukawa}
+\Delta \cL_{d > 4}~,
\ee
where $\Delta \cL_{d > 4}$ denotes the series of higher-dimensional 
operators invariant under the ST gauge group.
The coefficients of these operators have the form $c_i/\Lambda^{(d_i-4)}$,
where $c_i$ is an adimensional coefficient and $d_i$ denotes the canonical 
dimension of the effective operator.
If the new dynamics appears at the TeV scale, as we expect from a natural stabilization 
of the mechanism of electroweak symmetry breaking, the scale 
$\Lambda$ cannot exceed a few TeV.  Moreover, from naturalness arguments,\cite{'tHooft:1979bh} 
we should also expect that all the adimensional coefficients $c_i$ are of 
$O(1)$ unless suppressed by some symmetry argument.
The observation that this expectation is {\em not} 
fulfilled by several dimension-six operators contributing 
to flavour-changing processes is often denoted 
as the {\em flavour problem}.

If the ST Lagrangian were invariant under some flavour symmetry, 
this problem could easily be circumvented. 
For instance in the case of  baryon- or lepton-number violating 
processes, which are exact symmetries of the ST Lagrangian,
we can avoid the tight experimental bounds promoting 
$B$ and $L$ to be exact symmetries of the new dynamics 
at the TeV scale. The peculiar aspects of flavour 
physics is that there is no exact flavour symmetry 
in the low-energy theory. In this case it is 
not sufficient to invoke a flavour symmetry
for the underlying dynamics. We also need to specify how this 
symmetry is broken in order to describe the observed low-energy 
spectrum and, at the same time, be in agreement 
with the precise experimental tests
of flavour-changing processes.

The best way to quantify the flavour problem is obtained 
by looking at consistency of the tree- and loop-mediated 
constraints on the CKM matrix.
In first approximation we can assume that New Physics (NP) effects are
negligible in processes which are dominated by tree-level 
amplitudes. Following this assumption, the values of 
$|V_{us}|$, $|V_{cb}|$, and $|V_{ub}|$, as well as the 
constraints on $\alpha$ and $\gamma$ 
can be considered as NP free. As can be seen in Fig.~\ref{fig:UT},
this implies we can determine completely the CKM matrix assuming generic 
NP effects in loop-mediated amplitudes.
We can then use the measurements of observables which are
loop-mediated within the ST to bound the couplings of 
effective higher-dimensional operators 
which contribute to these observables at the tree level.

\begin{figure}[t]
  \centering
  {\includegraphics[width=0.6\textwidth]{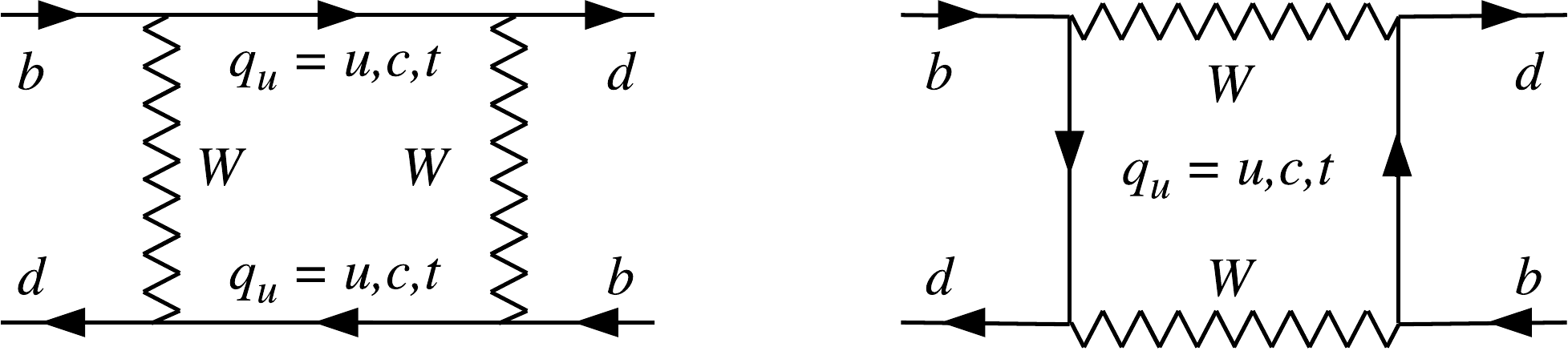}}
 \caption{Box diagrams contributing to $B_d$-$\bar B_d$ mixing in the unitary gauge.}
  \label{fig:BBmix}
\end{figure}

The loop-mediated constraints shown in Fig.~\ref{fig:UT}
are those from the mixing of $B_d$, $B_s$, and $K^0$
with the corresponding anti-particles
(generically denoted as $\Delta F=2$ amplitudes).
Within the ST, these processes are generated by 
box amplitudes of the type in Fig.~\ref{fig:BBmix}
(and similarly for $B_s$, and $K^0$) and are affected 
by small hadronic uncertainties. 
The leading contribution 
is obtained with the top-quark running 
inside the loop, giving rise to the highly suppressed 
result
\be
 \cM_{\Delta F=2}^{\rm ST} \approx  \frac{ G_F^2 m_t^2 }{16 \pi^2} ~ V_{3i}^* V_{3j} ~
    \langle \bar  M |  (\bar d_L^i \gamma^\mu d_L^j )^2  
| M \rangle \times  F\left(\frac{m_t^2}{m_W^2}\right)
\qquad [M = K^0, B_d, B_s]~,
\label{eq:DF2}
\ee
where $F$ is a loop function of  $O(1)$ and 
$i,j$ denote the flavour indexes of the meson
valence quarks.

Magnitude and phase of all these
mixing amplitudes have been determined 
with good accuracy from experiments and 
are consistent with the ST predictions. 
To translate this information into bounds on the scale 
of new physics, let's consider the following set of 
$\Delta F=2$ dimension-six operators in $\Delta \cL_{d > 4}$:
\be
\Delta \cL_{d > 4} \supset  \sum ~\frac{c_{ij}}{\Lambda^2} ~
\cO_{\Delta F=2}^{ij} \qquad   \cO_{\Delta F=2}^{ij} = (\bar q_L^i \gamma^\mu q_L^j )^2 ~.
\label{eq:dfops}
\ee
These operators contribute at the tree-level to 
the meson-antimeson mixing amplitudes. 
The condition 
$| \cM_{\Delta F=2}^{\rm NP}| <  | \cM_{\Delta F=2}^{\rm ST} |$
implies
\bea
\Lambda < \frac{ 3.4~{\rm TeV} }{| V_{3i}^* V_{3j}|/|c_{ij}|^{1/2}  }
<  \left\{ \ba{l}  
9\times 10^3~{\rm TeV} \times |c_{21}|^{1/2} \qquad {\rm from} \quad 
K^0-\bar K^0
 \\
4\times 10^2~{\rm TeV} \times |c_{31}|^{1/2} \qquad {\rm from}  \quad
B_d-\bar B_d
 \\
7\times 10^1~{\rm TeV} \times |c_{32}|^{1/2} \qquad {\rm from}  \quad
B_s-\bar B_s \ea
\right. 
\label{eq:boundsDF2}
\eea

\begin{table}[t]
\footnotesize{
\begin{center}
\begin{tabular}{c|c c|c c|c} \hline\hline
\rule{0pt}{1.2em}%
Operator &  \multicolumn{2}{c|}{$\Lambda$~in~TeV~($c_{\rm NP}=1$)} &
\multicolumn{2}{c|}{Bounds on
$c_{\rm NP}$~($\Lambda=1$~TeV) }& Observables\cr
&   Re& Im & Re & Im \cr  
 \hline $(\bar s_L \gamma^\mu d_L )^2$  &~$9.8 \times 10^{2}$& $1.6 \times 10^{4}$ 
&$9.0 \times 10^{-7}$& $3.4 \times 10^{-9}$ & $\Delta m_K$; $\epsilon_K$ \\ 
($\bar s_R\, d_L)(\bar s_L d_R$)   & $1.8 \times 10^{4}$& $3.2 \times 10^{5}$ 
&$6.9 \times 10^{-9}$& \  $2.6 \times 10^{-11}$ &  $\Delta m_K$; $\epsilon_K$ \\ 
 \hline $(\bar c_L \gamma^\mu u_L )^2$  &$1.2 \times 10^{3}$& $2.9 \times 10^{3}$ 
&$5.6 \times 10^{-7}$& $1.0 \times 10^{-7}$ & $\Delta m_D$; $|q/p|, \phi_D$ \\ 
($\bar c_R\, u_L)(\bar c_L u_R$)   & $6.2 \times 10^{3}$& $1.5 \times 10^{4}$ 
&$5.7 \times 10^{-8}$& $1.1 \times 10^{-8}$ &  $\Delta m_D$; $|q/p|, \phi_D$\\ 
\hline$(\bar b_L \gamma^\mu d_L )^2$    &  $6.6 \times 10^{2}$ & $ 9.3 \times 10^{2}$ 
&  $2.3 \times 10^{-6}$ &
$1.1 \times 10^{-6}$ & $\Delta m_{B_d}$; $S_{\psi K_S}$  \\ 
($\bar b_R\, d_L)(\bar b_L d_R)$  &   $  2.5 \times 10^{3}$ & $ 3.6
\times 10^{3}$ &  $ 3.9 \times 10^{-7}$ &   $ 1.9 \times 10^{-7}$ 
&   $\Delta m_{B_d}$; $S_{\psi K_S}$ \\
\hline $(\bar b_L \gamma^\mu s_L )^2$    &  $1.4 \times 10^{2}$ &  $  2.5 \times 10^{2}$   &  
 $5.0 \times 10^{-5}$ &   $1.7 \times 10^{-5}$ 
   & $\Delta m_{B_s}$; $S_{\psi \phi}$ \\ 
($\bar b_R \,s_L)(\bar b_L s_R)$  &    $ 4.8  \times 10^{2}$ &  $ 8.3  \times 10^{2}$  & 
   $8.8 \times 10^{-6}$ &   $2.9 \times 10^{-6}$  
  & $\Delta m_{B_s}$;  $S_{\psi \phi}$ \\ \hline\hline
\end{tabular}
\end{center}
}
\begin{tabnote} 
Table~1: Bounds on representative dimension-six $\Delta F=2$  
operators with effective coupling $c_{\rm NP}/\Lambda^2$.
The bounds are quoted on $\Lambda$, setting 
$|c_{\rm NP}|=1$, or on  $c_{\rm NP}$, setting
$\Lambda=1$ TeV. The right column denotes the main observables used to derive 
these bounds\cite{NirRev}.  
\end{tabnote} 
\end{table}

A more refined analysis, 
with complete statistical treatment and separate bounds 
for the real and the imaginary parts of the various amplitudes,  considering 
also operators  with different Dirac structure, leads to the bounds reported in Table~1.
The  main message of these bounds is the following:
\begin{itemize}
\item New physics models with a generic flavour 
structure ($c_{ij}$ of order 1) at the TeV scale are
ruled out. If we want 
to keep $\Lambda$ in the TeV range, physics beyond the ST
must have a highly non-generic flavour structure.  
\end{itemize}
In the specific
case of the $\Delta F=2$ operators in  (\ref{eq:dfops}),
in order to  keep $\Lambda$ in the TeV range,
we must find a symmetry argument such that  
$|c_{ij}| \lsim  |V_{3i}^* V_{3j}|^2$.
Reproducing a similar structure beyond the ST is a highly
non-trivial task. However, as discussed below, it can be 
obtained under specific assumptions.

\section{The Minimal Flavour Violation hypothesis}

The ``protection'' of  $\Delta F=2$ observables and, more generally,
flavour-changing  neutral-current (FCNC) processes occurring within the ST is a 
consequences of the specific symmetry and symmetry-breaking structure 
of the ST Lagrangian discussed in section~\ref{sect:ST}.
In particular, the fact that the quark flavour group $\cG_q$
is broken only by the two quark Yukawa couplings, and that the top-quark 
Yukawa coupling is the only $O(1)$ entry in $Y_{u,d}$,
is the main reason why Eq.~(\ref{eq:DF2}) is highly suppressed.
 
The strongest assumption we can make to suppress flavour-changing effects in generic 
extensions of the ST is the so-called Minimal Flavour Violation (MFV) 
hypothesis, namely the assumption that $Y_u$ and $Y_d$ are the only sources of flavour symmetry 
breaking also  beyond the ST.\cite{Chivukula:1987py,Hall:1990ac,D'Ambrosio:2002ex} 
 To implement and interpret this hypothesis in a consistent way, we can assume that $\cG_q$ is a good symmetry and promote $Y_{u,d}$ to be non-dynamical fields (spurions) with non-trivial transformation properties under $\cG_q$:
\begin{equation}
Y_u \sim (3, \bar 3, 1)\,,\qquad
Y_d \sim (3, 1, \bar 3)\,.
\end{equation}  
Employing the EFT language, an effective theory satisfies the MFV criterion in the quark sector if all higher-dimensional operators, constructed from ST 
fields and the $Y_{u,d}$ spurions, are formally invariant under the flavour group $\cG_q$~\cite{D'Ambrosio:2002ex}. 
The dynamical idea behind this construction is the hypothesis that the  breaking of the symmetry occurs at very high energy scales, 
and that $Y_{u,d}$  are the only independent combination of breaking terms (e.g.~combination of appropriate vacuum expectation values) 
that survive at low energies. 

According to the MFV criterion one should in principle consider operators with arbitrary powers of the (dimensionless) Yukawa fields. However, a strong simplification arises by the observation that all the eigenvalues of the Yukawa matrices are small, but for the top-quark one, and that the off-diagonal elements of the CKM matrix are very suppressed. This fact is enough to ensure that, even when including high powers of $Y_u$ and $Y_d$,
 FCNC amplitudes get exactly the same CKM suppression as in the ST:
\be
 \cM_{\Delta F=1}^{\rm MFV} (d^i \to d^j)  \propto  (V^*_{ti} V_{tj})~, \qquad 
 \cM_{\Delta F=2}^{\rm MFV} ( d^i \bar d^j  \to d^j  \bar d^i  )  \propto  (V^*_{ti} V_{tj})^2~.
\label{eq:FC}
\ee
The proportionality constants in these relations are flavour universal, implying 
the same NP correction (relative to the ST) in $s\to d$, $b\to d$, and  $b\to s$ transitions.

As a consequence of this structure, within the MFV framework several of the constraints used to determine the 
CKM matrix (and in particularly the unitarity triangle in Fig.~\ref{fig:UT}) are not affected by NP.\cite{Buras:2000dm}
For instance, the structure of the basic flavour-changing 
coupling in Eq.~(\ref{eq:FC}) implies that the weak CPV phase of 
$B_d$--$\bar B_d$ mixing is arg[$(V_{td}V_{tb}^*)^2$],
exactly as in the ST. 
This construction thus provides a
natural (a posteriori) justification of why no NP effects have 
been observed in the quark sector.
Moreover, the built-in CKM suppression
leads to bounds on the effective scale of new physics 
in the few TeV domain . These bounds are very similar to the 
bounds on flavour-conserving operators derived by precision electroweak tests. 

A few additional comments about the MFV hypothesis are listed below: 
\begin{itemize}
\item{}
Although MFV seems to be a natural solution to the flavour problem, we are still far from having proved the validity of this hypothesis from data. A proof of the MFV hypothesis can be achieved only with a positive evidence of NP exhibiting the flavour-universality pattern
predicted by MFV (same relative correction in $s\to d$, $b\to d$, and $b\to s$ transitions of the same type). 
This could happens, for instance, via precise measurements of the rare decays 
 $B_s\to\mu^+\mu^-$ and $B_d\to\mu^+\mu^-$.\cite{Bsmm_LHCb_new,Bsmm_CMS_new,Bsmm_combination}
Conversely, an evidence of NP in 
flavour-changing transitions not respecting the MFV pattern (e.g.~an evidence of $\cB(B_d\to\mu^+\mu^-)$
 well above its ST prediction) would not only imply the existence of physics beyond the ST, 
but also the existence of new sources of flavour symmetry breaking beyond the Yukawa couplings.
\item{}
The MFV ansatz is quite successful on the phenomenological side; however, 
it is unlikely to be an exact property of the model valid to all energy scales. 
Despite some recent attempts to provide a dynamical justification 
of this symmetry-breaking ansatz, the most natural possibility 
is that MFV is only an accidental low-energy  property of the theory.
It could also well  be that a less minimal connection between NP flavour-violating 
couplings and Yukawa couplings is at work. It is then very important to search for possible deviations (even if tiny) 
from the MFV predictions.
\item{}
Even if the MFV ansatz holds, it does not necessarily 
imply small deviations from the ST predictions in all flavour-changing phenomena. 
The MFV ansatz can be implemented in different ways. For instance, in models with 
two Higgs doublets we can change the relative normalization of the two Yukawa 
couplings.~\cite{D'Ambrosio:2002ex} It is also possible to decouple the breaking of CP invariance 
from the breaking of the $\cG_q$
quark-flavour group~\cite{Kagan:2009bn}, leaving more room for NP in CP-violating 
observables. All these variations lead to different and well defined patterns 
of possible deviations from the ST that we have only started to investigate
and that represent one of the main goal of present and future experiments in 
flavour physics.\cite{Bediaga:2012py,Abe:2010gxa,fortheNA62:2013jsa,
Yamanaka:2012yma}

\item{}
The usefulness of the MFV ansatz is closely linked to the theoretical expectation of NP in the TeV range.
This expectation follows from a  {\em natural} stabilization of the Higgs sector, but it is in tension with 
the lack of any direct signal of NP at the LHC. The more the scale of NP is pushed up, the more it is possible to 
allow sizable deviations from the MFV ansatz.
\end{itemize}

\section{Flavour symmetry breaking beyond MFV}

As anticipated,  MFV is not the only option to ``protect" flavour-changing transitions in extensions of the ST.
A key feature common to most models able to accommodate NP not far from the TeV scale, ensuring a sufficient 
suppression of flavour-changing transitions, 
is some link between flavour-changing amplitudes and fermion masses. 
Indeed the strong phenomenological bounds on flavour-changing transitions always 
involve light quarks (or leptons) of the first two generations, and are particularly strong 
in the case of transitions among the first two families (see Table~1).
Given the smallness of fermion masses of the first two generations, 
a link between  flavour-changing  amplitudes and fermion masses
provides a good starting point for a natural suppression of 
flavour-changing transitions. 

In the quark sector  this link can be efficiently implemented considering only  the $U(2)^3$ subgroup of the 
full quark  flavour group ($\cG_q$) that is obtained in the limit of vanishing Yukawa couplings for the first 
two generations of quarks.\cite{Barbieri:2011ci,Kagan:2009bn}
This symmetry limit is a better approximation 
of the full ST Lagrangian, since top and bottom quarks are allowed to have a 
non-vanishing mass.  The $U(2)^3$ subgroup is also sufficient to ensure enough
protection from flavour-changing transitions beyond the ST, assuming  
the minimal breaking structure necessary to describe light fermion masses.
The main difference of this ansatz compared to the MFV hypothesis is the breaking 
of the universal link between  $s\to d$ transitions vs.~transitions involving third
generation quarks ($b\to d$ and $b\to s$).
 
So far we discussed mainly the quark sector, but a flavour problem exists also in the lepton sector.
Similarly to the $\Delta F=2$ bounds in Table~1, the strong experimental bounds on 
FCNC transitions of charged leptons ($\mu\to e\gamma$, $\mu\to 3e$,  $\mu N\to e N$,
$\tau \to \mu\gamma$, \ldots) can be translated into bounds on NP scales 
well above the TeV, for $O(1)$ flavour-changing coefficients. 
For instance the MEG bound\cite{Adam:2013mnn}
$\cB(\mu\to e\gamma) < 5.7 \times 10^{-13}$  leads to an effective bound on $\Lambda$
of the order of $10^5$~TeV.  

In order to allow  TeV scale NP, some extension of the  MFV hypothesis can be implemented also  is the lepton sector.  
However, given there is not a unique way to accommodate non-vanishing neutrino masses, in this case  
there is more freedom  to define the minimal sources of flavour symmetry breaking. Different versions of Minimal Lepton Flavour 
Violation (MLFV) have been proposed in the literature, depending on how the irreducible breaking terms in the neutrino sector are 
identifed~\cite{Cirigliano:2005ck,Grinstein:2006cg,Davidson:2006bd,Gavela:2009cd,Alonso:2011jd}. 
On general grounds, it is not difficult to provide a sufficient suppression of  flavour-changing coefficients 
for TeV scale new physics, provided the (adimensional) flavour breaking terms associated to neutrino masses are sufficiently small.
In the context of see-saw models, this imply masses for the heavy right-handed neutrinos typically around or 
below $10^{12}$~GeV.\cite{Cirigliano:2005ck} A significant progress in this field is expected by the next generation 
of LFV experiments with charged leptons, especially in the sector of $\mu\to e$ transitions.\cite{Kuno:2015tya}
As for the quark sector,  the key tool to test flavour symmetries 
 (and symmetry-breaking) assumptions relies on the observation of
  possible correlations in the rate of neutral-current LFV processes, such as $\tau\to \mu\gamma$ vs.~$\mu\to e\gamma$.

\section{Flavor physics and partial compositeness}
 
In the previous two sections we have discussed mechanisms to suppress flavour-changing
transitions beyond the ST due to specific  flavour symmetries and  
symmetry-breaking patterns. An interesting 
alternative is the possibility of a generic {\em  dynamical suppression}
of flavour-changing interactions, related to the weak mixing of 
the light ST fermions with some new dynamics occurring at the TeV scale.
This is what happens in the so-called models with partial compositeness,
\cite{Kaplan:1991dc,Agashe:2004cp}  where 
the hierarchy of fermion masses 
is attributed to the hierarchical mixing of the ST fermions with 
the heavier (composite) states of the theory.

Also the general features of this class 
of models can be described by means of an effective theory approach.\cite{Davidson:2007si,KerenZur:2012fr}
The two main assumptions of this EFT approach are
the following: 
\begin{itemize}
\item{}
There exists a (non-canonical) basis 
for the ST fermions where their kinetic terms exhibit 
a rather hierarchical form: 
\bea
&& {\cal L}^{\rm quarks}_{\rm kin} = \sum_{\Psi=Q_L, U_R, D_R }
 \overline{\Psi} Z_\psi^{-2}   D\sla \Psi~, \no\\
&& Z_\psi =  {\rm diag}(z_\psi^{(1)}, z_\psi^{(2)}, z_\psi^{(3)}  )~, 
\qquad  z_\psi^{(1)}\ll z_\psi^{(2)}\ll z_\psi^{(3)} \lsim 1~.
\eea
\item{} In such basis there  is no flavour symmetry and all the 
flavour-violating interactions, including the Yukawa 
couplings, are $\cO(1)$.  
\end{itemize}
Once the fields are transformed 
into the canonical basis, the hierarchical kinetic terms 
act as a distorting lens, through which all interactions 
are seen as approximately aligned on the magnification axes of the lens. 
The hierarchical $z_\psi^{(i)}$ can be interpreted as the effect 
of the mixing of an elementary  (ST-like)  sector of massless fermions 
with a corresponding set of heavy composite fermions: the elementary fermions feel 
the breaking of the electroweak (and flavour) symmetry only via this mixing. 

The values of the $z_\psi^{(i)}$ can be deduced,
up to an overall normalization, from the know structure of the
Yukawa couplings, that can be decomposed as follows
\be
Y_u^{ij} \sim z_Q^{(i)}  z_U^{(j)}~, \qquad 
Y_d^{ij} \sim z_Q^{(i)}  z_D^{(j)}~.
\ee 
Inverting such relations we can express the $z_\psi^{(i)}$ combinations 
appearing in the effective couplings of dimension-six operators involving ST fields
[e.g.~the combination $(z_Q^{(1)} z_Q^{(2)})^2$ for the operator $(\bar s_L \gamma_\mu d_L)^2$, etc\ldots]
into appropriate powers of quark masses and CKM angles. The resulting suppression 
of FCNC amplitudes turns out to be quite effective 
being linked to the hierarchical structure of the Yukawa couplings. 

As shown in a recent analysis,~\cite{KerenZur:2012fr} this framework is compatible 
with the strong flavour bounds in kaon sector for scales of the composite states (vector resonances)
around $10$ TeV. In this case  one can expect deviations from the ST at the present level of experimental sensitivity 
 in the  electric dipole moment (EDM) of the neutron (where there is actually a significant tension with the present bound),  
 CP-violating observables in the kaon system ($\epsilon'/\epsilon$ and $\epsilon_K$),
 and $b\to s$ FCNC transitions. However, in the lepton sector
 the minimal framework is not satisfactory (a severe fine-tuning is needed to satisfy current bounds on lepton-flavour violating processes).

It should be stressed that also in  partial-compositeness models is possible to postulate the existence of additional 
protective flavour symmetries (as discussed e.g.~in Ref.~\refcite{Redi:2011zi,Barbieri:2012bh,Barbieri:2012tu}) and, for instance, recover a MFV structure. 
In this case the bounds  on the composite states turn out to be well below $10$ TeV.
     
\section{Dynamical Yukawa couplings}

The  MFV principle does not provide an explanation for  the observed pattern of  masses and mixings of quarks and leptons:
the Yukawa couplings are simply treated as inputs, as in the ST. To a large extent, also the mechanism of partial compositeness 
does not explain the observed pattern of quark and lepton Yukawa couplings: the hierarchal mixing between elementary and composite 
fermions is an input of the construction.

A more ambitious goal is that of deriving the observed structure of the Yukawa couplings from some fundamental principle.
The simplest realization of the idea of a dynamical character for the Yukawa couplings is to assume that
 \be
 {\it Y}=\frac{\langle 0|\Phi|0\rangle }{\Lambda}
 \label{vev1}
 \ee
with $\Lambda$ being some high energy scale and $\Phi$ a set of scalar fields (or composite operators) with transformation properties such as to make invariant the effective  potential $V({\it Y})$ under the flavour group $\cal G_{\rm flavour}$ (or some of its subgroups). 
A general problem that one encounters along this line 
is the  unwanted  appearance of a large number of Goldstone bosons, associated to the spontaneous breaking 
of the large global continuos flavour symmetry. This problem could be avoided assuming that the 
 flavour symmetry is gauged at some high energy scale.~\cite{Grinstein:2010ve}

An interesting alternative to continuos flavour symmetries, that naturally avoids the problem of Goldstone bosons,
is the possibility that the fundamental flavour symmetry is a suitable discrete subgroup of $\cal G_{\rm flavour}$.
This option has received a lot of attention in the recent past, mainly because of neutrino physics:\cite{Altarelli:2010gt}
the neutrino mixing matrix exhibits an almost tri-bimaximal structure and the latter 
is naturally expected in the context of discrete flavour symmetries. 
However, the description of both quark and lepton sectors in terms of a unique discrete flavour symmetry
is less trivial and significantly more complicated\cite{Altarelli:2010gt,Feruglio:2015jfa,Zwicky:2009vt}.
Moreover, this option has become less appealing also in the pure neutrino sector  
after the observation of a sizable $1$--$3$ neutrino mixing angle,\cite{Ahn:2012nd,An:2012eh}
that implies sizable deviations from the tri-bimaximal mixing structure.

The idea that quark masses and, more generally, the Yukawa couplings, could arise from the minimization of a potential invariant under some 
continuos flavour symmetry is an old idea. Earlier attempts dates back to the sixties, when  Michel and Radicati~\cite{Michel:1970mua}, 
and Cabibbo and Maiani~\cite{Cabibbo:1970rza}  developed generic group-theoretical methods to identify  the {\em natural extrema} 
of $SU(3)_L\times SU(3)_R $ invariant potentials. Several further attempts towards a dynamical origin of the Yukawa couplings, employing 
various subgroups of $\cal G_{\rm flavour}$ have been discussed in the literature~\cite{Froggatt:1978nt,Ibanez:1994ig,Anselm:1996jm, Barbieri:1999km,Berezhiani:2001mh,Harrison:2005dj,Feldmann:2009dc,Alonso:2011yg,Nardi:2011st,Alonso:2012fy,Espinosa:2012uu}. 
 In models based on small symmetry groups, such as the $U(1)$ horizontal symmetry originally  proposed by Froggat and Nielsen~\cite{Froggatt:1978nt}, it is quite easy to reproduce the observed mass matrices in terms of
a reduced number of free parameters, while it is difficult to avoid problems with  FCNCs, unless some amount of 
fine-tuning is introduced. 

In models based on large (MFV-like) symmetry groups, it is difficult to explain the full pattern of quark and lepton masses 
in absence of significant fine-tuning among the coefficients of the potential.\cite{Alonso:2011yg}
In this context, an interesting recent development has been presented  in Ref.~\refcite{Alonso:2013nca}.
There it has been shown that, among the most stable solution of the general 
minimization problem of $V({\it Y})$, corresponding to maximally unbroken subgroups\cite{Michel:1970mua,Cabibbo:1970rza}
of $\cal G_{\rm flavour}$, there exists a class of solutions quite close to a realistic spectrum. 
In the quark sector, this corresponds to a  hierarchical mass pattern of the third vs.~the first two generations, with unity  CKM matrix. 
In the lepton sector, it implies hierarchical masses for charged leptons and degenerate Majorana neutrinos, 
with one maximal, one large, and one vanishing mixing angle. Both these textures are close to the real situation, and can be brought in full agreement with data adding small perturbations.
In the neutrino sector, this implies a firm prediction that can be tested in the near future, namely an almost degenerate spectrum with an average
neutrino mass close to $m_\nu \approx  0.1~{\rm eV}$.

The radical alternative to predictions of quark and lepton masses based on continuos or discrete symmetries is the idea 
that they are simply random variables, possibly selected by anthropic arguments. The latter option has recently 
gained consensus, given the lack of deviations from the ST after the first run of the LHC.\cite{Giudice:2013yca}  
Drawing any firm conclusion in this respect is very difficult, and it will remain so also in the future. 
However, it is worth to stress that  the measurement of the absolute value of neutrino masses could 
provide a significant additional piece of this fascinating puzzle: a value close to the present bounds, compatible with 
the hypothesis of a degenerate spectrum, would certainly speak in favour of some underlying large and 
mildly broken flavour symmetry.\cite{Alonso:2013nca,Blankenburg:2012nx}

\section{Conclusions}

Flavour physics has a twofold role in investigating the nature of physics beyond the ST. 
On the one hand, for NP models with new particles close to the TeV scale, existing low-energy 
flavour-physics measurements put very stringent limits on the flavour structure of the model.
As illustrated in general terms and with a few specific examples, for such models 
present data tell us that the new degrees of freedom must have a highly non-generic
flavour structure.  However, this structure has not been clearly identified yet.   
In this perspective, if direct signals of NP will appear during the next LHC run, 
future progress in flavour physics will be an essential  tool to  investigate 
the peculiar flavour structure of the new degrees of freedom.

On the other hand, the paradigm of NP at the TeV scale is seriously challenged by the absence of 
deviations from the SM at the high-energy frontier. In this perspective, flavour physics remains a very powerful
tool to search for physics beyond the ST, being potentially sensitive to NP scales much higher than 
those directly accessible at present and near-future high-energy facilities. 

\bigskip

{\footnotesize 
\noindent 
{\bf Acknowledgements}
I thank Admir Greljo, Luciano Maiani, and David Marzocca for useful comments on the manuscript.
This work was supported in part by the Swiss National Science Foundation (SNF) under contract 200021-159720.}

\end{document}